\documentclass[amsmath,showpacs,groupedaddress,nofootinbib,superscriptaddress,10pt,twocolumn]{revtex4}
\usepackage{amssymb}
\usepackage{graphicx}
\usepackage{dcolumn}
\usepackage{bm}
\usepackage{graphicx}
\usepackage{epsfig}
\usepackage{color}
\usepackage{graphics}
\usepackage{amsthm}
\usepackage{amsthm}
\usepackage{dcolumn}
\usepackage{float}

\usepackage{ulem}
\begin{document}
\title{Holographic energy density on Ho\v{r}ava-Lifshitz cosmology}
\author{Samuel Lepe}
\email[]{slepe@ucv.cl} \affiliation{Instituto de F\'\i sica,
Facultad de Ciencias, Pontificia Universidad Cat\'olica de
Valpara\'\i so, Avenida Brasil 4950, Valpara\'\i so, Chile.\\}
\author{Francisco Pe\~{n}a}
\email[]{francisco.pena@ufrontera.cl} \affiliation{Departamento de
Ciencias F\'isicas, Facultad de Ingenier\'ia y Ciencias,
Universidad de La Frontera, Casilla 54-D, Temuco, Chile.}
\author{Francisco Torres}
\email[]{francisco.t2@gmail.com} \affiliation{Instituto de F\'\i
sica, Facultad de Ciencias, Pontificia Universidad Cat\'olica de
Valpara\'\i so, Avenida Brasil 4950, Valpara\'\i so, Chile.\\}
%\date{\today}
\begin{abstract}
In Ho\v{r}ava-Lifshitz cosmology we use the holographic Ricci-like
cut-off for the energy density proposed by L. N. Granda and A.
Oliveros and under this framework we study, through the cosmic
evolution at late times, the sign change in the amount of
non-conservation energy ($Q$) present in this cosmology. We revise
the early stage (curvature-dependent) of this cosmology, where a
term reminiscent of stiff matter is the dominant, and in this
stage we find a power-law solution for the cosmic scale factor
although $\omega =-1$. Late and early phantom schemes are obtained
without requiring $\omega <-1$. Nevertheless, these schemes are
not feasible according to what is shown in this paper. We also
show that $ \omega =-1$ alone does not imply a de Sitter phase in
the present cosmology. Thermal aspects are revised by considering
the energy interchange between the bulk and the spacetime boundary
and we conclude that there is no thermal equilibrium between them.
Finally, a ghost scalar graviton (extra degree of freedom in HL
gravity) is required by the observational data.
\end{abstract}

\maketitle

\section{Introduction}

The Ho\v{r}ava-Lifshitz (HL) cosmology \cite{Mukohyama2010} is a
formalism generated from HL gravity (a possible candidate for
quantum gravity?), which is a power-counting renormalizable
gravity theory (expected to be renormalizable and unitary) that
leads to modifications of Einstein's general relativity at high
energies producing novel features for considering cosmology. This
formalism suffers from the lack of local Hamiltonian constraint
and thus there is no a Friedmann equation here. Therefore, if a
projectability condition is imposed (the lapse function is
restricted  being only time-dependent), then the Hamiltonian
constraint becomes one global. This means that the Hamiltonian
constraint in HL gravity is not a local equation but an equation
integrated over a whole space and the projectability condition is
compatible with the foliation preserving diffeomorphism
(diffeomorphism invariance). In HL gravity, in addition to the
tensor graviton, the theory exhibits an extra scalar degree of
freedom called {\it scalar graviton} and, as we shall see, the
role of this extra degree of freedom (which we "characterize" by a
$\eta $-parameter) will play an important role alongside
parameters from the observational data.

In the present work we follow the philosophy developed in
\cite{Kobayashi2009}, Section III. In particular, the claim cited
there "{\it{the global Hamiltonian constraint that}} $Q$ (amount
of non-conservation energy) {\it{does not necessarily vanish in
the local patch}}" is the main key of our work. In
\cite{Mukohyama2010} and under this scope, it is shown that $Q$ is
not zero today and only today we have $C_{0}/a^{3}$, that is,
$\epsilon \left( t\right) =const/a^{3}$(according to the notation
of \cite{Kobayashi2009} and \cite{MukohyamaWangMaartens2009}).

In HL cosmology (described here in a
Friedmann-Lemaitre-Robertson-Walker universe) we have a
non-conservation equation for the cosmic fluid, and the sign
change experienced by the amount of non-conservation energy
through the cosmic evolution, treated in the present paper as the
energy interchange between the bulk and the spacetime boundary and
we mean by bulk the observable universe and by boundary its Hubble
horizon. The thermal equilibrium between them will be discussed by
using a holographic cut-off for the energy density and we will
discuss also early and late phantom solutions obtained and its
factibility in the present cosmology.

The paper is organized as follows: Considering the flat case in
Section II, we inspect the sign change in the amount of
non-conservation energy through the cosmic evolution by using the
$q$-parameter (deceleration parameter) and a phantom solution is
found without requiring $\omega <-1$. In Section III we use a
$q(z)$-parametrization in order to visualize the sign change of
the aforementioned $Q(z)$. In Section IV we use a
$\omega(z)$-parametrization in order to complement the discussion
on the sign change of $Q(z)$ done from Section III. In Section V
we study the early limit of HL cosmology and we find a scheme with
$\omega =-1$ where the Hubble parameter exhibits a power-law
behavior and we find a phantom scheme in which also $\omega =-1$.
In Section VI we analyze some thermal aspects under the idea of a
sign change of $Q(z)$. Section VII is devoted to presents our
conclusions. $G=c=1$ units will be used.

\section{ Holographic Ricci-like on flat Ho\v{r}ava-Lifshitz cosmology}

We consider the dynamic equation \cite{Mukohyama2010}
\begin{eqnarray}  \label{eq1}\eta (2\dot{H}+3H^{2})=-\omega \rho ,
\end{eqnarray}
where $\eta =(3\lambda -1)/2$ with $\lambda $ being a
dimensionless parameter fixed by the diffeomorphism invariance of
4D general relativity (GR) and $0<\eta <1$ (ghost instability;
ghost scalar graviton), $\eta <0$ or $\eta >1$ (non-ghost scalar
graviton) and $\eta $ is fixed to $1$ in GR. $H$ is the Hubble
parameter, $\omega =p/\rho $ being $p$ the pressure and $\rho $
the energy density and the dot means the temporal derivative. The
non-conservation equation for the energy density is given by
\begin{eqnarray} \label{eq2}\dot{\rho}+3H\left( 1+\omega \right) \rho =-Q,
\end{eqnarray}
where $Q$ is the amount of non-conservation energy present in the
model and the low energy limit can be recovered if
$Q\longrightarrow 0$. We recall here that $Q$ comes from the
theory (as an integration term in HL-cosmology) and not imposed by
hand when, for example, interacting fluids are treated. Now, by
considering $\rho $ as a dominant component (the unspoken
components, if any, are negligible), we will interpret the sign
change of $Q$ as energy transference between the bulk and the
spacetime boundary.

Although under a different scope at the present work, in
\cite{Xiao-Dong2013} we see observational evidence for $Q\left(
0\right) >0$, namely decay of dark energy into dark matter. From
our perspective, we can affirm something similar, that is, decay
of energy from the bulk into the boundary of spacetime.

By introducing in (\ref{eq1}) the holographic energy density model
\cite{GrandaOliveros2008}, written in terms of the\ $q$-parameter
defined by $q=-\left( 1+\dot{H}/H^{2}\right) $,
\begin{eqnarray}  \label{eq3}\rho =3\left( \alpha H^{2}+
\beta \dot{H}\right) =3\left[ \alpha -\beta \left( 1+q\right)
\right] H^{2},
\end{eqnarray}
where $\alpha $ and $\beta $ are both positive dimensionless
constant parameters \cite{Lepe2010}, we obtain
\begin{eqnarray}  \label{eq4}\eta \left[ 3-2\left( 1+q\right) \right] =
-3\omega \left[\alpha -\beta \left( 1+q\right) \right] .
\end{eqnarray}
And from this last expression, in addition to (\ref{eq2}), it is
straightforward to obtain
\begin{eqnarray}  \label{eq5}\frac{Q}{3H^{3}}=-2\eta \left[1-\frac{\alpha}
{\eta}+\frac{\beta}{\eta}\left(1+q\right)\right]\left(q-\frac{1}{2}\right)
\nonumber\\ -\beta \left(1+z\right)\frac{dq}{dz},
\end{eqnarray}
where we have used the redshift parameter defined by $1+z=a_{0}/a$
being $a$ the cosmic scale factor. The expression given in
(\ref{eq5}) will be the central key if we are consider the sign
change of $Q$ through the evolution \cite{BingWang2007,
Arevalo2014}. By considering $\beta =0$, we write
\begin{eqnarray}  \label{eq6}\frac{Q}{6H^{3}}=-\eta
\left( 1-\frac{\alpha }{\eta }\right) \left( q-\frac{1
}{2}\right),
\end{eqnarray}
and we can already visualize an explicit sign change of $Q$, for
fixed $sgn(1-\alpha /\eta)$, given the sign change of $q(z)$ at
some time during the evolution. Additionally, from (\ref{eq5}) we
can see that $Q(q=1/2) \neq 0$ and $Q=Q(q^{2}) $, a quadratic
dependence on $q$, and this fact is fully $\beta $-dependent
\cite{Arevalo2014}. According to (\ref{eq6}), $ \alpha =\eta $ or
$q=1/2$, both lead to $Q=0$. In Section III we will discuss with
more detail the expressions (\ref{eq5}-\ref{eq6}) by introducing a
$q$ -parametrization while taking into account the inequality
$\alpha /\eta \gtrless 1$. For instance, if $0<\alpha /\eta <1$
and given that $q(0)<0$, from (\ref{eq6}) we have $Q(z=0) >0$ and
we have energy transference today from the bulk to the boundary.

We discuss briefly some differences between GR and the present
holographic HL cosmology. With (\ref{eq4}) we write
\begin{eqnarray}\label{eq7}
 1+q=\frac{3}{2}\left(\frac{1+
\frac{\alpha}{\eta}\omega
}{1+\frac{3\beta}{2\eta}\omega}\right)\,\,\, and \,\,\,
q-\frac{1}{2}=\frac{3}{2\eta }\nonumber \\
\times \left(\alpha -\frac{3\beta}{2}\right) \left(\frac{\omega
}{1+\frac{3\beta} {2\eta}\omega}\right) \, ,
\end{eqnarray}
and if we do $\beta =0$ we have
\begin{eqnarray}\label{eq8}
1+q=\frac{3}{2}\left( 1+\frac{\alpha } {\eta }\omega \right) \,
\,\,\, and \,\,\, q-\frac{1}{2}=\frac{3}{2}\frac{\alpha }{\eta
}\omega  \, ,
\end{eqnarray}
and in GR
\begin{eqnarray} \label{eq9}
1+q=\frac{3}{2}\left( 1+\omega \right) \, \,\,\, and \,\,\,
q-\frac{1}{ 2}=\frac{3}{2}\omega  \, ,
\end{eqnarray}
and we note that in both cases (HL and GR) $\omega =0$ leads to
$q=1/2$. So, $\alpha /\eta $ and $\beta /\eta $ do the difference.

We examine now some cases by considering $\omega =const $. In this
case, the solutions for $H(t)$, in addition to the cosmic scale
factor $a(t)$, are, respectively,
\begin{eqnarray}\label{eq10}
H(t)=H_{0}\left[1+\frac{3}{2}\left( \frac{1+\frac{
\alpha}{\eta}\omega }{1+\frac{3\beta}{2\eta}\omega }\right)
H_{0}\left( t-t_{0}\right)\right]^{-1},
\end{eqnarray}
\begin{eqnarray}\label{eq11} a(t)=a_{0}
\left[1+\frac{3}{2}\left( \frac{1+\frac{ \alpha}{\eta}\omega
}{1+\frac{3\beta}{2\eta}\omega }\right) H_{0}\left(
t-t_{0}\right)\right] ^{\frac{ 2}{3} \Theta},
\end{eqnarray}
where $\Theta =\left[1+\left(3\beta /2\eta \right) \omega \right]
/\left[1+\left( \alpha /\eta \right) \omega \right]$, and we
recall that $0<\eta <1 $: the scalar graviton is a ghost and
otherwise (no ghost) if $\eta >1$. As we have just seen, $\eta <0$
is discarded: from (\ref{eq8}) we have $q(0)$ and $\omega(0)$ both
negatives and so $\eta >0$. If we consider $\omega <0$, we write
(\ref{eq10}-\ref{eq11}) in the form
\[
H(t)= H_{0}\left[1+\frac{3}{2}\left(\frac{1-\frac{\alpha}{\eta}
\left\vert \omega \right\vert}{1-\frac{3\beta}{2\eta}\left\vert
\omega \right\vert}\right) H_{0}\left(t-t_{0}\right)\right] ^{-1},
\]
\begin{eqnarray}\label{eq12}
a(t)=a_{0}\left[1+\frac{3}{2}\left(\frac{1-\frac{\alpha}{\eta}
\left\vert \omega \right\vert}{1-\frac{3\beta}{2\eta}\left\vert
\omega \right\vert}\right) H_{0}\left(t-t_{0}\right)\right]
^{-\frac{2}{3}\left\vert \Theta \right\vert },
\end{eqnarray}
and a phantom scheme arises if $3\beta /2 \eta\left\vert \omega
\right\vert <1<(\alpha / \eta)\left\vert \omega \right\vert$ which
is consistent with $\alpha / \beta >3/2$, (see
\cite{GrandaOliveros2008, Arevalo2014})
\begin{eqnarray}\label{eq13}H(t)=\frac{2}{3}\left\vert
\frac{1-\frac{3\beta}{2\eta}\left\vert\omega \right\vert
}{1-\frac{\alpha}{\eta}\left\vert\omega\right\vert}\right\vert
\left(t_{s}-t\right)^{-1}\;\; and \nonumber\\
t_{s}=t_{0}+\frac{2}{3}\left\vert\frac{1-\frac{3\beta}{2\eta}
\left\vert\omega\right\vert}{1-\frac{\alpha}{\eta}\left\vert
\omega\right\vert}\right\vert H_{0}^{-1},
\end{eqnarray}
and we have a phantom evolution without requiring $\omega <-1$.
Given that $ H(t\rightarrow t_{s})\rightarrow \infty $,
$a(t\rightarrow t_{s}) \rightarrow \infty $, $\rho(t\rightarrow
t_{s}) \rightarrow \infty $ and $p\rightarrow \infty $ , the
present singularity is Type I (Big Rip) \cite{Nojiri2005}. From
(\ref{eq10}), a de Sitter phase can be obtained if we do $(\alpha
/\eta)\omega =-1$ and $\alpha /\eta \neq 1$. So, in the present
scheme, $\omega =-1$ alone does not imply a de Sitter evolution.
On the other hand, the inequality given after (\ref{eq12}) tells
us that we can confine $\eta $ to the range $3\beta /2<\eta
<\alpha $, given that $\left\vert \omega(0) \right\vert \sim 1$.
Additionally, in line with to (\ref{eq13}) and (\ref{eq6}),
$Q(t\longrightarrow t_{s}) \longrightarrow \infty $. While it is
true that the above inequality, $3\beta /2<\eta <\alpha $, is
consistent with $\eta
>3\beta /2$, it is not consistent with $\alpha <\eta $, if $Q(0) >0$
(see the next Section). So, if we do not want a phantom scheme, we
must have $\alpha /\eta <1$ and then $H(z)$ is well behaved (free
of singularities). The fact, $\alpha /\eta <1$, could be an
antecedent to consider for deleting the phantom scheme given in
(\ref{eq13}).

Finally, we revise the limit $\eta \rightarrow \infty $
\cite{Gumrukcuoglu2011}. If we want a finite $Q$, in this limit
according to (\ref{eq5})) and (\ref{eq6}), we must have $q=1/2$,
so that $Q=-\beta (1+z) dq/dz\neq 0$ at $z=\bar{z}\neq \infty $
(see next Section) and $Q=0$, respectively. From (\ref{eq7}) and
(\ref{eq8}) we have the same: $q=1/2$. From (\ref{eq10}) the
solution for the Hubble parameter is the same as GR, that is, an
evolution driven by dust: $H(t)=H_{0}\left[1+\left(3/2\right)
H_{0}\left( t-t_{0}\right)\right] ^{-1}$ and the phantom solution
disappears given that it is not possible to satisfy the inequality
$1<\left( \alpha /\eta \right) \left\vert \omega \right\vert $,
when $\eta \rightarrow \infty $.

\section{$Q(z)$ and $q(z)$\;-\;parametrization}

In order to have the best visualization of the sign change of
$Q(z)$, we will use the $q$-parametrization given by
\begin{eqnarray}\label{eq14} q(z)-\frac{1}{2}=q_{1}
\frac{(z+q_{2}/q_{1})}{(1+z)^{2}},
\end{eqnarray}
where $q_{1}=1.47_{-1.82}^{+1.89}$, $q_{2}=-\left(1.46\mp
0.43\right) <0$ \cite{GongWang2007}, and we verify that
\begin{eqnarray}  \label{eq15}q(z=-q_{2}/q_{1}) =
q(\infty) =\frac{1}{2},
\end{eqnarray}
i. e., there are two values of $z$ for which $q=1/2$ and this fact
will be relevant in the following. By doing $q_{2}=-\left\vert
q_{2}\right\vert $ and $\overline{z}=\left\vert q_{2}\right\vert
/q_{1}$, we write (\ref{eq14}) in the equivalent form
\begin{eqnarray}\label{eq16}q(z) -\frac{1}{2}=
q_{1}\frac{(z-\overline{z})}{( 1+z) ^{2}},
\end{eqnarray}
so that the derivative on $q(z)$ reads
\begin{eqnarray}\label{eq17}\frac{dq(z)}{dz}=q_{1}
\left(1+2\overline{z}\right)\left(1-
\frac{z}{1+2\overline{z}}\right)\left(1+z\right)^{-3},
\end{eqnarray}
and
\begin{eqnarray}\label{eq18}\frac{dq(z)}{dz}
\left\vert _{q=1/2}\right. =\frac{dq\left( z\right)
}{dz}\left\vert _{z=\overline{z}}\right. =\frac{q_{1}}{\left(1+
\overline{z}\right)^{2}}>0.
\end{eqnarray}
By doing $\left\vert q_{2}^{\left(+\right)}\right\vert =1.03$,
$\left\vert q_{2}^{\left(-\right)}\right\vert =1.83$,
$q_{1}^{\left(+\right)}=3.36$,\;\; $q_{1}^{\left(-\right)}=-0.35$,
we write
\begin{eqnarray}\label{eq19}\begin{array}{cc}
\vert q_{2}^{(+)}\vert /q_{1}^{(+)}=0.31, & \vert q_{2}^{(+)
}\vert /q_{1}^{(-)}=-2.94, \\
\vert q_{2}^{(-)}\vert /q_{1}^{(+)}=0.54, & \vert q_{2}^{(-)}\vert
/q_{1}^{(-)}=-5.23,
\end{array}
\end{eqnarray}
and we must discard $\vert q_{2}^{(+)}\vert /q_{1}^{(-)}$ and
$\vert q_{2}^{(-)}\vert /q_{1}^{(-)}$(in both cases,
$\overline{z}<-1$). In other words, $q_{2}=-(1.46\mp 0.43)$ and
$q_{1}=3.36$ in order to be consistent. So, $\overline{z}=0.31$ or
$\overline{z}=0.54$, both lie in the past, and the derivative in
(\ref{eq18}) is positive.

We come back now to (\ref{eq17}). Given that $q_{1}>0$ and
$\overline{z}>0$, we see that the sign change on the derivative
occurs at $z=1+2\overline{z}$ ($=1.62$ or $2.08$). See, for
instance, $q(\overline{z}) =1/2$ and $ dq(z) /dz=0$ at
$z=1+2\overline{z}>\overline{z}$, and do not forget also that
$q(\infty) =1/2$. Additionally, $q(0) =1/2-$ $q_{1}\overline{z}<0$
($q(0)=-0.54$, if $ \overline{z}=0.31$ and $q(0) =$ $-1.31$, if
$\overline{z}=0.54$) and $dq(0) /dz=q_{1}(1+2\overline{z})
>0$. So, if we consider the option $q\left( \infty \right) =1/2$,
from (\ref{eq17}) we have $dq\left( z\right) /dz=0$ at $z=\infty
$, and this result is not good if we want to see the sign change
of $Q(z)$ (see (\ref{eq5})). So, for the present parametrization
we must choose $q(\overline{z})=1/2$ and not $q(\infty) =1/2$. The
sign change of the acceleration occurs, roughly, at
$\widetilde{z}\sim 0.7$ \cite{Magaña2014}, such that hereinafter
we will use  $\overline{z}=0.54$ by considering that this value is
closer to $ \widetilde{z}$.

Now, by replacing (\ref{eq16}) in (\ref{eq6}), where $\beta =0$,
we can write
\begin{eqnarray}\label{eq20}\frac{Q(z)}{6H^{3}}=
-\eta \left( 1-\frac{\alpha }{\eta }\right)
q_{1}\frac{(z-\overline{z}) }{(1+z) ^{2}},
\end{eqnarray}
and we have a clear sign change of $Q$, i. e., $sgnQ(z) =sgn(z-
\overline{z})$ given that $sgn(1-\alpha /\eta) $ is provided as
fixed. If we use (\ref{eq16}) and (\ref{eq17}) in (\ref{eq5}), we
can write
\[
\left(1+z\right) ^{2}\frac{Q(z)} {3\eta
q_{1}H^{3}}=-2\left[1-\frac{\alpha }{\eta }+\frac{\beta
}{\eta}\left(\frac{3}{2}+q_{1}\frac{\left(z-\overline{z}\right)}
{\left(1+z\right)^{2}}\right)\right]
\]
\begin{eqnarray} \label{eq21}
\times
\left(z-\overline{z}\right)-\frac{\beta}{\eta}
\left(1+2\overline{z}\right)
\left(1-\frac{z}{1+2\overline{z}}\right).
\end{eqnarray}
Now, if we consider $q_{1}=3.36$ and $\overline{z}=0.54$ we have
$Q(0)>0$ if $0<\alpha /\eta <1$ and $0<\beta /\eta <0.45$ (see
\cite{Lepe2010, Arevalo2014}), and it is straightforward to show
that $Q(z\rightarrow \infty ) <0$ and $Q( z\rightarrow -1) <0$,
that is,we have a double sign change of $Q(z)$ if $\beta \neq 0$.
This fact is
clearly shown in Figure 1.\\
\begin{figure}[!ht]
\begin{center}
\includegraphics[width=6cm]{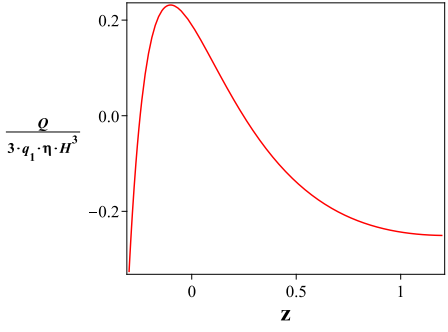}
\caption{\label{Figure1}The\ picture shows the double sign change
that $ Q\left( q\right)$ exhibits through the evolution. We
observe a one change in the past and the other in the future. Here
$q_{1}=3.36$, $\overline{z}=0.54$, $\alpha /\eta =0.6$\ and $\beta
/\eta =0.1$. So, we have today energy transference from the bulk
to the spacetime boundary.}
\end{center}
\end{figure}
\\
From $q(z)$ to $\omega(z)$, the quotient $\alpha /\eta $ and the
$\eta $-value. By replacing (\ref{eq16}) into (\ref{eq4}) we
obtain
\[
\omega(z)=\frac{2\eta
q_{1}}{3}\left[\frac{z-\overline{z}}{\left(\alpha -3\beta
/2\right)\left(1+z\right)^{2}-\beta
q_{1}\left(z-\overline{z}\right)}\right] ,
\]
and from here we can have a feeling of the $\eta $-value
\[
\eta
=-\frac{3\omega\left(0\right)}{2q_{1}\overline{z}}\left(\alpha
-\frac{3\beta}{2}+\beta q_{1}\overline{z}\right),
\]
given that we know $\omega(0)$, $q_{1}$, $\overline{z}$, $ \alpha
$ and $\beta $. For instance, if we use $q_{1}=3.36$,
$\overline{z} =0.54$ and the set of values for $\omega(0)$,
$\alpha$ and $\beta$ given in \cite{Lepe2010}, we can verify that
$\alpha /\eta <1$ and $0<\eta <1$ (ghost graviton), and so we
verify also that $Q(0)>0$.
\section{ $Q(z)$ and $\omega(z)$\;-\;parametrization}

We come back to (\ref{eq1}) and we solve by incorporating
(\ref{eq3}) and by doing $\beta =0$ , for simplicity. First, we
consider $\omega=const$. In this case, the solution of
(\ref{eq1}) is
\begin{eqnarray}
\label{eq22} H\left( z\right) =H_{0}
\left( 1+z\right) ^{3\left( 1+\left( \alpha /\eta
\right) \omega \right) /2},
\end{eqnarray}
where we have used the redshift parameter $z$ defined by
$1+z=a_{0}/a$, where $a$ is the cosmic scale factor. By using
(\ref{eq3}) in (\ref{eq2}) we obtain
\begin{eqnarray}  \label{eq23}\frac{Q}{9\alpha H^{3}}\left( z\right) =-\omega
\left( 1-\frac{\alpha }{\eta
}\right) ,
\end{eqnarray}
where $H(z)$ is given in (\ref{eq22}). Using (\ref{eq4}) we can
write, for instance,
\begin{eqnarray}\label{eq24}q(0)=\frac{1}{2}\left[1+3\frac{\alpha}
{\eta}\omega(0)\right],
\end{eqnarray}
where $\omega(0)<0$ and $q(0)<0$ are both observational
parameters, such that we are able to write the following
constraint for $\alpha /\eta $
\begin{eqnarray}\label{eq25}\frac{\alpha}{\eta}>\frac{1}{3}\left\vert \omega
\left( 0\right) \right\vert \sim \frac{1}{3},
\end{eqnarray}
then $\eta>0$, as has been seen before. Therefore, according to
(\ref{eq23}) and (\ref{eq25}) we obtain
\begin{eqnarray}\label{eq26}\frac{Q}{9\alpha H^{3}}\left(0\right) =
-\omega(0)\left(1- \frac{\alpha }{\eta }\right) =\left\vert
\omega(0) \right\vert \left(1-\frac{\alpha}{\eta
}\right)\nonumber\\ \Longrightarrow Q(0)
>0 \;\; if \;\;\frac{\alpha}{\eta}<1,
\end{eqnarray}
and then, if we want $Q(0) >0$, we can establish the following
constraint for the ratio $\alpha /\eta$ : $1/3<\alpha /\eta <1$.\

If later on we have $\omega <0$ and we consider (\ref{eq22}), we
have
\[
1-\left( \alpha /\eta \right)\left\vert \omega \right\vert =
\]
\begin{equation}\label{eq27}
\left\{
\begin{array}{c}
>0\longleftrightarrow \left\vert \omega \right\vert
<(\frac{\alpha}{\eta}
)^{-1}\Longrightarrow H(z\rightarrow -1) \rightarrow 0:
qe,\\
<0\longleftrightarrow \left\vert \omega \right\vert
>(\frac{\alpha}{\eta})^{-1}\Longrightarrow H(z\rightarrow -1)
\rightarrow \infty :ph,
\end{array}
\right.
\end{equation}
and if $\alpha/\eta=\left\vert \omega \right\vert ^{-1}<1$
($\left\vert \omega \right\vert >1$, like phantom), we have $H(z)
=const$, i.e., a de Sitter phase and not necessarily $\omega =-1$.

We inspect now (\ref{eq22}) and (\ref{eq23}) by considering
different stages of the evolution (each characterized by $\omega
=const$). According to (\ref{eq26}) and independently of $\eta $,
for $\omega =0$ (dust as the dominant component) we have $Q=0$.
For $\omega >0$  we will always have $sgnQ(z) =-sgn(
1-\alpha/\eta) <0$ and if $\omega <0$, $sgnQ(z) =sgn(
1-\alpha/\eta) >0$, if $\alpha/\eta <1$. And yet there nothing
that we can visualize yet about the possibility of an explicit
sign change of $Q(z)$ through the evolution if $\omega =const$. We
study this possibility by using the usual
Chevallier-Polarski-Linder parametrization given by
$\omega(z)=\omega(0)+{\omega}^{\prime}(0)z(1+z)^{-1}$
\cite{Chevallier2001} and as we can see in the literature
\cite{Jassal2005}, the sign of ${\omega}^{\prime}(0)=$ $d\omega
/dz$ at $z=0$, is model-dependent according to the fit-values from
the observational data. In our analysis, however, we discuss both
options for the sign of ${\omega}^{\prime}(0)$. Therefore, using
(\ref{eq1}) and (\ref{eq3}), it is straightforward to obtain the
following solution for the Hubble parameter
\begin{eqnarray}\label{eq28}H(z) =H_{0}
\left(1+z\right)^{\left( 3/2\right)\left[1+\left(\alpha /\eta
\right)\left(\omega\left(0\right)+{\omega}^{\prime}\left(0\right)
\right)\right]}\nonumber\\\times \exp \left(
-\frac{3}{2}\frac{\alpha}{\eta}{\omega}^{\prime}\left(0\right)
\frac{z}{1+z}\right) ,
\end{eqnarray}
and in this case we obtain for $Q(z)$ the expression
\begin{eqnarray}\label{eq29}
\frac{Q\left( z\right) }{9\alpha H_{0}^{3}} =-\omega(0)\left(
1+z\right)^{\left( 9/2\right) \left[ 1+\left( \alpha /\eta \right)
\left({\omega}^{\prime}(0) +{\omega}(0)\right)\right] }\nonumber
\\\times \left[ 1-\frac{\alpha }{\eta
}+\frac{{\omega}^{\prime}(0)}{\omega(0)}\frac{z}{1+z}\right] \exp
\left(-\frac{9}{2}\frac{\alpha}{\eta}{\omega}^{\prime}(0)
\frac{z}{1+z}\right) ,
\end{eqnarray}
and
\[
\frac{Q(z\rightarrow \infty)}{9\alpha
H_{0}^{3}}\longrightarrow -\omega(0)\left( 1+z\right)^{\left(
9/2\right) \left[ 1+\left( \alpha /\eta \right) \left(\omega(0)
+{\omega}^{\prime}(0)\right) \right] }
\]
\begin{eqnarray}\label{eq30}
\times \left[1-\frac{\alpha}{\eta}+\frac{{\omega}^{\prime}(0)}
{\omega(0)}\right] \exp \left( -\frac{9}{2}\frac{\alpha }{\eta
}{\omega}^{\prime}(0)\right) ,
\end{eqnarray}
where $sgn\left[Q(z\rightarrow \infty)\right] =sgn\left[1-\alpha
/\eta +{\omega}^{\prime}\left(0\right) /\omega \left( 0\right)
\right]$. If $Q\left( z\rightarrow \infty \right) $ has to be
finite and positive at early times (reasonable consideration, see
[1]), we must have $ 1+\left( \alpha /\eta \right) \left( \omega
\left( 0\right) +{\omega}^{\prime} \left( 0\right) \right) <0$,
i.e., ${\omega}^{\prime}\left( 0\right) <0$, given that
$\left\vert \omega \left( 0\right) \right\vert \sim 1$ and
$1-\alpha /\eta +{\omega}^{\prime}\left( 0\right) /\omega \left(
0\right) >0$.

Now, we inspect now the sign change of $Q\left( z\right) $. By
using (\ref{eq29}), we perform
\begin{eqnarray}  \label{eq31}Q\left( z_{0}\right) =0
\longleftrightarrow z_{0}=-\left( 1+\frac{{ \omega}^{\prime}\left(
0\right) /\omega \left( 0\right) }{1-\alpha /\eta }\right) ^{-1},
\end{eqnarray}
and then $-1<z_{0}<0$ if $\alpha /\eta <1$ and
${\omega}^{\prime}\left( 0\right) <0$, given that $\omega \left(
0\right) <0$. So, $Q\left( z\right) $ will undergo a sign change
in the future. This fact is consistent with the display of Figure
1. For completeness, by using (\ref{eq31}) we write (\ref{eq29})
in the form
\[
\frac{Q\left( z\right) }{9\alpha H_{0}^{3}} =-\omega \left(
0\right) \left( 1-\frac{\alpha }{\eta }\right) \left( 1+z\right)
^{\left( 9/2\right) \left[ 1+\left( \alpha /\eta \right) \left(
\omega \left( 0\right) +{ \omega}^{\prime}\left( 0\right) \right)
\right] }
\]
\begin{equation}\label{eq32}
\times \left[ 1-\left( \frac{1+z_{0}}{1+z} \right)
\frac{z}{z_{0}}\right] e^{-\frac{9}{2}\frac{\alpha }{\eta
}{\omega}^{\prime}\left( 0\right) \frac{z}{1+z}} ,
\end{equation}
and so we can better visualize the sign change of $Q\left(
z\right) $. We note that in (\ref{eq32}), the potency of $\left(
1+z\right) ^{9/2}$ can be written in the form
\begin{eqnarray}  \label{eq33}1-\left( \alpha /\eta \right)
\left\vert \omega \left( 0\right) \right\vert
\left[ 1+\left\vert {\omega}^{\prime}\left( 0\right) /\omega
\left( 0\right) \right\vert \right] ,
\end{eqnarray}
and, independently of $sgn(1-(\alpha /\eta) \vert
\omega(0)\vert (1+ \vert {\omega}^{\prime}(0)
/\omega(0)  \vert ))$, the exponential behavior is what
decides the treated limits.

We consider now the limit $Q\left( z\longrightarrow -1\right) $.
In this case, according to (\ref{eq29}) we have
\begin{eqnarray}  \label{eq34}\frac{Q\left( z\longrightarrow -1\right) }
{9\alpha H_{0}^{3}}\longrightarrow {\omega}^{\prime}\left(
0\right)\nonumber\\
\times \lim_{z\longrightarrow -1}\left( 1+z\right) ^{\left(
9/2\right) \left[ 7/9+\left( \alpha /\eta \right) \left( \omega
\left( 0\right) +{\omega}^{\prime}\left( 0\right) \right) \right]
}\nonumber
\\ \times \exp \left( \frac{9}{2}\frac{\alpha }{\eta }{\omega}^{\prime}\left(
0\right) \frac{1}{1+z} \right) ,
\end{eqnarray}
and if ${\omega}^{\prime}\left( 0\right) <0$, then $Q\left(
z\longrightarrow -1\right) \rightarrow -0$ and, according to
(\ref{eq28}), $H$ $\left( z\rightarrow -1\right) \rightarrow 0$.
Therefore, there is energy transference from the bulk to the
boundary as just was seen in Section III.
\begin{figure}[h!]
\begin{center}
\includegraphics[width=6cm]{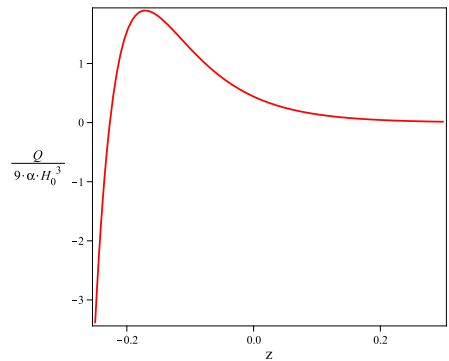}
\caption{The picture shows only a sign change of $Q\left( z\right)
$\ (in the future) due to the fact that we considered $\beta =0$.
If we compare with Figure 1, we note clearly the role of $\beta
\neq 0$. Here $ \alpha /\eta =0.6$, $\omega \left( 0\right)
=-1.1$\ and $\acute{\omega} \left( 0\right) =-1.5$}
\label{Figure2}
\end{center}
\end{figure}

Finally, we have to ask what happens if we consider
${\omega}^{\prime}\left( 0\right)
>0$? This
is an open issue to discuss (do not forget that $sgn\left[
{\omega}^{\prime} \left( 0\right) \right] $ is model-dependent,
see \cite{Jassal2005}). For the purposes of this discussion, for
instance, we can maintain the idea of $Q\left( 0\right) >0$ and
thus discuss the consistency with the previous analysis done in
this Section. This discussion will not be undertaken here.

\section{$ \rho =3 \left( \alpha {H}^{2}+\beta
\dot{H}\right)$  on early Ho\v{r}ava-Lifshitz cosmology}

We consider now the following equations
\begin{eqnarray}  \label{eq35}3\eta H^{2}=\rho -\gamma \frac{k^{3}}{a^{6}},
\end{eqnarray}
and
\begin{eqnarray}  \label{eq36}\eta \left( 2\dot{H}+3H^{2}\right) =-\omega \rho +
\gamma \frac{k^{3}}{a^{6}},
\end{eqnarray}
where $k=0,\pm 1$ (curvature index) and $\gamma $ is a constant.
In Ho\v{r}ava-Lifshitz cosmology, the scheme (35-36) is generated
by considering only terms in which $a^{-6\text{ }}$ is the
dominant one \cite{Mukohyama2010}. A term like $\gamma
k^{3}a^{-6}$ is reminiscent of stiff matter (in GR: $p=\rho
\rightarrow \rho \sim a^{-6}$) and its presence (stiff matter) at
early stages of the evolution may have played an important role
under the scope of a holographic approach to cosmology
\cite{BanksFischler04}. In \cite{Bambi2014} a Friedmann equation
is derived from a four-fermion interaction, and where a
$a^{-6}$-like term appears naturally which is used there, in
addition to a dust term $a^{-3}$, to avoid possible cosmic
singularities. From (\ref{eq35}, \ref{eq36}) we can obtain
\begin{eqnarray}  \label{eq37}2\eta \left( \dot{H}+3H^{2}\right) =
\left( 1-\omega \right) \rho ,
\end{eqnarray}
and after replacing the given cut-off in the last equation, we
have
\begin{eqnarray}  \label{eq38}d\ln H\left( z\right) =
\left( \frac{2-\left( \alpha /\eta \right) \left( 1-\omega \right)
}{2-\left( 3\beta /\eta \right) \left( 1-\omega \right) } \right)
d\ln \left( 1+z\right) ^{3},
\end{eqnarray}
and if $\omega =1$ or $\eta \longrightarrow \infty $, or also
$\alpha =3\beta $ (although this equality is not required by
observational fits \cite{Lepe2010, Arevalo2014}), the Hubble
parameter and the cosmic scale factor are, respectively,
\[
H\left( z\right) =H_{0}\left( 1+z\right) ^{3}\longrightarrow
\]
\begin{eqnarray}  \label{eq39}
a\left( t\right) =a_{0}\left[ 1+3H_{0}\left( t-t_{0}\right)
\right] ^{1/3}\Longrightarrow \ddot{a}\left( t\right) <0,
\end{eqnarray}
i.e., the same solution as found in GR ($3H^{2}=\rho $ and
$\dot{\rho}+3H\left( 1+\omega \right) \rho =0$ with $\omega =1$);
nevertheless, in the present case we have non-null curvature (see
(\ref{eq35}, \ref{eq36})). If we consider now $\omega =-1 $, we
have
\[
H\left( z\right) =H_{0}\left( 1+z\right) ^{3\Delta }\text{ \ \ }
\]
and\text{ \ \ \ }
\begin{eqnarray}  \label{eq40}
\Delta =\left( 1-\alpha /\eta \right) /\left( 1-3\beta /\eta
\right) ,
\end{eqnarray}
and the cosmic scale factor and the acceleration are given,
respectively, by
\begin{eqnarray}  \label{eq41}a\left( t\right) =a_{0}\left[ 1+
3\Delta H_{0}\left( t-t_{0}\right) \right]
^{1/3\Delta },
\end{eqnarray}
(a power-law solution) and
\begin{eqnarray}  \label{eq42}\ddot{a}\left( t\right) =
a_{0}H_{0}^{2}\left( 1-3\Delta \right) \left[
1+3\Delta H_{0}\left( t-t_{0}\right) \right] ^{\left( 1/3\Delta
\right) -2}.
\end{eqnarray}
Now, if $1/6<\Delta <1/3$ the acceleration given in (\ref{eq42})
is positive and decreases in time although $\omega =-1$. But, at
early times we expect to have something like $H\sim H_{0}$ (old
inflation-like), i. e., according to (\ref{eq40}) $\alpha /\eta
\sim 1\Longrightarrow \omega \sim -1$ and in this case $a\left(
t\right) \sim \exp \left( H_{0}t\right) $. Hence, the ratio
$\alpha /\eta $ is very important here. In particular, if we put
$\beta =0$ ($ \Delta =1-\alpha /\eta
>0\longleftrightarrow \alpha /\eta <1$) we write \ \
\begin{eqnarray}  \label{eq43}a\left( t\right) =a_{0}
\left[ 1+3\left( 1-\frac{\alpha }{\eta }\right) H_{0}\left(
t-t_{0}\right) \right] ^{1/3\left( 1-\alpha /\eta \right)
},\nonumber \\\Longrightarrow
H(t)=H_{0}\left[1+3\left(
1-\frac{\alpha }{\eta }\right) H_{0}\left(
t-t_{0}\right)\right]^{-1}
\end{eqnarray}
and the acceleration is
\begin{eqnarray}  \label{eq44}\ddot{a}\left( t\right) =
a_{0}H_{0}^{2}\left[ 1-3\left( 1-\frac{\alpha }{
\eta }\right) \right] \nonumber \\\times\left[ 1+3\left(
1-\frac{\alpha }{\eta }\right) H_{0}\left( t-t_{0}\right) \right]
^{-2\left( \alpha /\eta -5/6\right) /\left( \alpha /\eta -1\right)
},
\end{eqnarray}
and by considering the inequality $2/3<\alpha /\eta <5/6$ (and
$Q\left( z=0\right) >0$, see Sections II, III and IV) the
acceleration given in (\ref{eq44}) is , as was stipulated,
positive and decreases with time, although $\omega \approx -1$.
\begin{figure}[h!]
\begin{center}
\includegraphics[width=6cm]{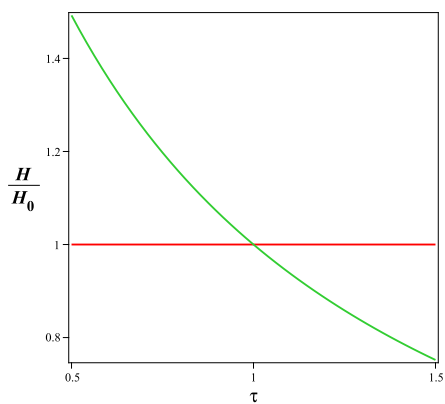}
\caption{The Hubble parameter given in (\ref{eq43}) compared with
$ H=const $(usual de Sitter phase). Here $\alpha /\eta =0.78$.}
\label{Figure3}
\end{center}
\end{figure}

A similar pattern to that shown in Figure 3 is discussed in
\cite{DiMarco2006} under the idea of "graceful" old inflation.
There, the authors have shown that cite: {\it a false vacuum can
successfully decay to a true vacuum, producing inflation, ...
,since exponential inflation is slowed down to power-law
Inflation}. Thus, the present scheme built under the holographic
philosophy of HL cosmology could be considered a new antecedent
giving an alternative to previous studies done.

Early phantom phase. We consider $\Delta <0$ ($3\beta <\eta
<\alpha $ or $ \alpha <\eta <3\beta $) (see [3,5] for $\alpha $
and $\beta $ values). In this case we have for the Hubble
parameter the solution $H\left( t\right) =\left( H_{0}/3\left\vert
\Delta \right\vert \right) \left( t_{s}-t\right) ^{-1}$, where
$t_{s}=t_{0}+\left( 1/3\left\vert \Delta \right\vert \right)
H_{0}^{-1}$ and for the scale factor we obtain $a\left( t\right)
=a_{0}\left( 3\left\vert \Delta \right\vert H_{0}\right)
^{-1/3\left\vert \Delta \right\vert }\left( t_{s}-t\right)
^{-1/3\left\vert \Delta \right\vert }$. If we put $\beta =0$ and
$\alpha /\eta >1$ ($\Delta <0$), the role of $\alpha /\eta $ in
the raised phantom scheme becomes  more evident. Do not forget
that we have with $\omega =-1$ from (\ref{eq40}) to (\ref{eq44}),
meaning that we have an "early" phantom scheme without requiring
$\omega <-1$. Nevertheless, at early times we do not have a
singularity according to the observational data and therefore, we
discard this early behavior and then we have again a strong
argument for claiming that $0<\alpha /\eta <1$ (and so, positive
$\eta $).

According with (\ref{eq42}), $\ddot{a}\left( t\right)
>0$ and not a phantom ($ \Delta >0$) $\Longrightarrow 0<\left(
1-\alpha /\eta \right) /\left( 1-3\beta /\eta \right) <1/3$, we
obtain
\begin{eqnarray}  \label{eq45}0<\left( \eta -
\alpha \right) /\left( \eta -3\beta \right)
<1/3\Longrightarrow \nonumber \\\alpha /\eta <1\text{ \ \
}and\text{ \ }\eta
>3\beta ,
\end{eqnarray}
or
\begin{eqnarray}  \label{eq46}0<\left( \alpha -\eta \right)/
\left( 3\beta -\eta \right)
<1/3\Longrightarrow \nonumber \\\alpha /\eta >1\text{ \ \
}and\text{ \ }\eta <3\beta ,
\end{eqnarray}
and in this last case we can have both options $\eta \gtrless 0$.
If $\eta <0 $, we write
\begin{eqnarray}  \label{eq47}0<\left( \left\vert \eta \right\vert +
\alpha \right) /\left( \left\vert \eta
\right\vert +3\beta \right) <1/3\Longrightarrow
\frac{2}{3}\left\vert \eta \right\vert +\alpha <\beta ,
\end{eqnarray}
and this option is discarded given that $\alpha >\beta $
\cite{Lepe2010, BingWang2007}. We note that, in the present
situation, the limit $\eta \rightarrow \infty $ does not operate.
So, $\alpha /\eta <1$ and we have no phantom and we agree with the
discussions performed before.

Finally, by using the $q$-parameter and the given holographic
cut-off, from (\ref{eq24}) we can write
\begin{eqnarray}  \label{eq48}1+q=3\left( \frac{2-\left( \alpha /\eta \right)
\left( 1-\omega \right) }{
2-3\left( \beta /\eta \right) \left( 1-\omega \right) }\right) ,
\end{eqnarray}
and we can notice that $\omega =1$ or $\eta \longrightarrow \infty
$, both leading to $q=2$ (usual stiff matter behavior as in GR).
If we are thinking in "old inflation", i.e., $\omega \sim -1$, we
have
\begin{eqnarray}  \label{eq49}1+q\sim 3\left( \frac{1-
\alpha /\eta }{1-3\beta /\eta }\right) ,
\end{eqnarray}
and only in the case $\alpha /\eta \sim 1$ we recover $q\sim -1$
(as GR).

\section{Thermal aspects}

Some studies have considered the possibility of energy interchange
between the bulk and the spacetime boundary \cite{LepePeña2014}.
If we are think about thermal equilibrium between the bulk and the
boundary, the answer is not reflecting our results. There is
non-equilibrium given that there are two changes in the direction
of the energy interchange: one in the past ($z>0$ ) and other in
the future ($-1<z<0$) if $\beta \neq 0$ or only one change (in the
future) if $\beta =0$. What is the sign of $Q\left( 0\right) $?
The answer is fully dependent on $\eta $, $sgn\left( \eta \right)
$ and the values of $\alpha $ and $\beta $. And we have $Q\left(
0\right) >0$ in accordance with the obtained and well justified
constraint $0<\alpha /\eta <1.$ On the other hand, if we consider
$\eta <0$ from heading and by using (\ref{eq6}), we have
\begin{eqnarray} \label{eq50}
\frac{Q}{6H^{3}}=\left\vert \eta \right\vert \left( 1+\frac{\alpha
}{ \left\vert \eta \right\vert }\right) \left(
q-\frac{1}{2}\right) ,
\end{eqnarray}
and given that $q\left( 0\right) <0$, then $Q\left( 0\right) <0$.
But, if we considerer(\ref{eq24}) we obtain
\begin{eqnarray} \label{eq51}
q\left( 0\right) =\frac{1}{2}\left( 1+3\alpha \left\vert
\frac{\omega \left( 0\right) }{\eta }\right\vert \right) >0,
\end{eqnarray}
and there is an inconsistency with the observational data:
$q\left( 0\right) <0$. Therefore, we apologize that $\eta >0$ is
consistent with the current observational data, and hence we
emphasize that $q\left( 0\right) <0$ is a crucial antecedent to
justify what we have developed here.

\section{Conclusions}

We have shown the existence of sign changes, through the
evolution, of the amount of non-conservation energy $Q\left(
z\right) $ present in the framework of the Ho\v{r}ava-Lifshitz
cosmology as a consequence of the philosophy of a holographic
scheme assigned to the energy-matter content in the theory. We
have analyzed the late limit and the early limit (where a
reminiscent stiff like-matter term is one dominant), and we do not
observe phantom phases, in any event, if $\alpha /\eta <1$. At
early times, we have found a power-law solution for the cosmic
scale factor, although $\omega =-1$, and this fact may be
considered a new antecedent to consider if we think about in old
inflation. Is it possible to respond to this concern, the old
inflation problem, within the framework of HL-cosmology under a
holographic scope? We leave this concern for now.

Additionally, $\eta $ has been relatively well confined to the
range $0<\eta <1$ according to the used observational setting for
$\alpha $, $\beta $ and $ \omega \left( 0\right) $. And so, we can
say we have a ghost scalar graviton in the present framework.

Finally, we live out the thermodynamic equilibrium and {\it{we are
cooling down}} ($Q\left( 0\right) >0$) according to what is shown
here (Figures 1 and 2).

\section*{Acknowledgments}
%\begin{acknowledgements}
This work was supported by Fondecyt Grant No.1110076 (S.L.) and
VRIEA-DI-PUCV Grant No.037.377/2014, Pontificia Universidad
Cat\'{o}lica de Valpara\'{\i}so (S.L.). (F.P.) acknowledges Grant
DI14-0007 of Direcci\'{o}n de Investigaci\'{o}n y Desarrollo,
Universidad de La Frontera.
%\end{acknowledgements}

\end{document}